\input lanlmac.tex
\overfullrule=0pt
\input epsf.tex
\figno=0
\def\fig#1#2#3{
\par\begingroup\parindent=0pt\leftskip=1cm\rightskip=1cm\parindent=0pt
\baselineskip=11pt
\global\advance\figno by 1
\midinsert
\epsfxsize=#3
\centerline{\epsfbox{#2}}
\vskip 12pt
{\bf Fig. \the\figno:} #1\par
\endinsert\endgroup\par
}
\def\figlabel#1{\xdef#1{\the\figno}}
\def\encadremath#1{\vbox{\hrule\hbox{\vrule\kern8pt\vbox{\kern8pt
\hbox{$\displaystyle #1$}\kern8pt}
\kern8pt\vrule}\hrule}}

%
\font\zfont = cmss10 
\font\litfont = cmr6

\def\bigone{\hbox{1\kern -.23em {\rm l}}}
\def\ZZ{\hbox{\zfont Z\kern-.4emZ}}
\def\hf{{\litfont {1 \over 2}}}

\def\p{\partial}
\def\a{\alpha}
\def\b{\beta}
\def\g{\gamma}
\def\d{\delta}

\def\u{\upsilon }

\def\vp{\varphi}

\def\L{\Lambda}

\def\o{\omega }


\def\hb{\hbar}
\def\bh{{1\over \hbar}}
\def\ibh{ {i\over \hb} }


   \def\CD {{\cal D}}
   
   \def\CF {{\cal F}}

   \def\CN {{\cal N}}

   \def\CR {{\cal R}}

   \def\CZ {{\cal Z}}

\def\bZ{ Z^\dagger  }

\def\bL{\bar L  }

   \def\R{\relax{\rm I\kern-.18em R}}
   \font\cmss=cmss10 \font\cmsss=cmss10 at 7pt
   \def\Z{\relax\ifmmode\mathchoice
   {\hbox{\cmss Z\kern-.4em Z}}{\hbox{\cmss Z\kern-.4em Z}}
   {\lower.9pt\hbox{\cmsss Z\kern-.4em Z}}
   {\lower1.2pt\hbox{\cmsss Z\kern-.4em Z}}\else{\cmss Z\kern-.4em
   Z}\fi}
 \def\p{\partial}
 
 \def\11{1\!\! 1}

 \def\({\left(}
 \def\){\right)}
 \def\[{\left[}
 \def\]{\right]}
 
\def\vvp{\Phi}
\def\bL{{\bar L}}

\def\bz{{\bar z}}
\def\Pe{\Psi^{_{E}}}
\def\pse{ \psi^{_{E}} }
\def\pee{\psi^{_{E'}}}


\def\xp{x_{_{+}}}
\def\xm{x_{_{-}}}
\def\xpm{x_{_{\pm}}}



\def\Xp{X_{+}}
\def\Xm{X_{-}}

\def\psip{\psi_{_{+}}}
\def\psim{\psi_{_{-}}}
\def\psipm{\psi_{_{\pm}}}

\def\Pepm{\Pe_{\pm }}

\def\psep{\pse_{_{+}}}
\def\psem{\pse_{_{-}}}
\def\psepm{\pse_{_{\pm }}}

\def\peep{\pee_{_{+}}}

\def\Pip#1{\Psi^{#1}_{+}}
\def\Pim#1{\Psi^{#1}_{-}}

\def\Phip{P_{+}}
\def\Phim{P_{-}}

\def\Tr{\,{\rm Tr}\,}

\def\Cb{{\rm \bf C}}
\def\Zb{{\rm \bf Z}}

\def\ctf{\sqrt{\Lambda} }
\def\ZM{\CZ}
\def\ZN{\CZ^{^{_{\rm NMM}}}}
\def\taum{\tau}
\def\taun{\tau^{^{_{\rm NMM}}}}


\chardef\tempcat=\the\catcode`\@ \catcode`\@=11
\def\cyracc{\def\u##1{\if \i##1\accent"24 i%
    \else \accent"24 ##1\fi }}
\newfam\cyrfam
\font\tencyr=wncyr10
\def\cyr{\fam\cyrfam\tencyr\cyracc}


\def\np#1#2#3{{\it Nucl. Phys.} {\bf B#1} (#2) #3}
\def\pl#1#2#3{{\it Phys. Lett.} {\bf B#1} (#2) #3}
\def\prl#1#2#3{{\it Phys. Rev. Lett.} {\bf #1} (#2) #3}
\def\physrev#1#2#3{{\it Phys. Rev.} {\bf D#1} (#2) #3}

\def\rmatp#1#2#3{{\it Rev. Math. Phys. }{\bf #1} (#2) #3}
\def\cmp#1#2#3{{\it Comm. Math. Phys.} {\bf #1} (#2) #3}
\def\mpl#1#2#3{{\it Mod. Phys. Lett.} {\bf #1} (#2) #3}
\def\ijmp#1#2#3{{\it Int. J. Mod. Phys.} {\bf #1} (#2) #3}
\def\lmp#1#2#3{{\it Lett. Math. Phys.} {\bf #1} (#2) #3}
\def\tmatp#1#2#3{{\it Theor. Math. Phys.} {\bf #1} (#2) #3}
\def\hepth#1{{\tt hep-th}/#1}

\lref\HKK{ J. Hoppe, V. Kazakov, and I. Kostov,
``Dimensionally reduced SYM$_4$ as solvable matrix quantum
mechanics'', \np{571}{2000}{479}, \hepth{9907058}.}
\lref\NKK{V. Kazakov, I. Kostov and  N.  Nekrasov,
``D-particles, matrix integrals and KP hierarchy'',
\np{557}{1999}{413},  \hepth{9810035}.}
\lref\KKK{V. Kazakov, I. Kostov, and D. Kutasov,
``A Matrix Model for the Two Dimensional Black Hole",
\np{622}{2002}{141}, \hepth{0101011}.}
\lref\AK{S. Alexandrov and V. Kazakov,
``Correlators in 2D string theory with vortex condensation'',
\np{610}{2001}{77}, \hepth{0104094}.}
\lref\AKTBH{S. Alexandrov and V. Kazakov, ``Thermodynamics of 2D string theory'',  
{\it JHEP} {\bf 0301} (2003) 078, \hepth{0210251}.}
\lref\IK{I. Kostov, ``String Equation for String Theory on a Circle'',
\np{624}{2002}{146}, \hepth{0107247}.}
\lref\AKK{S.Yu. Alexandrov, V.A. Kazakov, and I.K. Kostov,
``Time-dependent backgrounds of 2D string theory'',
\np{640}{2002}{119}, \hepth{0205079}.}
\lref\KM{V. Kazakov and A. Marshakov,
``Complex Curve of the Two Matrix Model and its Tau-function'',
\hepth{0211236}.}

\lref\KAZMIG{V.A. Kazakov and A.A. Migdal, ``Recent progress in the non-critical strings'' 
\np{311}{1988}{171}.}
\lref\BRKA{E. Brezin, V. Kazakov and Al. Zamolodchikov,
\np{338}{1990}{673}.}
\lref\PARISI{ G. Parisi, \pl{238}{1990}{209, 213}.}
\lref\GRMI{ D. Gross and N. Miljkovic, \pl{238}{1990}{217}.}
\lref\GIZI{P. Ginsparg and J. Zinn-Justin, \pl{240}{1990}{333}.}
\lref\BIPZ{ E. Brezin, C. Itzykson, G. Parisi, and J.-B. Zuber,
\cmp{59}{1978}{35}.}
\lref\DAVID{F. David, ``Non-Perturbative Effects in Matrix Models and
Vacua of Two Dimensional Gravity'',
\pl{302}{1993}{403}, \hepth{9212106}.}
\lref\DavidEy{ G. Bonnet, F. David, and B. Eynard, 
``Breakdown of universality in multi-cut matrix models'',
{\it J. Phys.} {\bf A33} (2000) 6739, {\tt cond-mat}/0003324.}
\lref\GRKL{D. Gross and I. Klebanov, \np{344}{1990}{475};
  \np{354}{1990}{459}.}
\lref\BULKA{D. Boulatov and V. Kazakov,
``One-Dimensional String Theory with Vortices as Upside-Down
Matrix Oscillator'', \ijmp{8}{1993}{809}, \hepth{0012228}. }
\lref\KAZ{V.A. Kazakov,
``Bosonic Strings and String Field Theories in One-dimensional
Target Space'' in ``Random Surfaces and Quantum Gravity'' ed. by
O. Alvarez, (1990) 269.}

\lref\DVV{R. Dijkgraaf, E. Verlinde, and H. Verlinde,
``String propagation in a black hole geometry'',
\np{371}{1992}{269}.}
\lref\FZZ{ V. Fateev, A. Zamolodchikov, and Al. Zamolodchikov,
{\it unpublished}.}

\lref\MP{G. Moore and M. Plesser,
``Classical scattering in 1+1 Dimensional string theory'',
\physrev{46}{1992}{1730}, \hepth{9203060}.}
\lref\MPR{G. Moore, M. Plesser, and S. Ramgoolam,
``Exact S-matrix for 2D string theory'',
\np{377}{1992}{143}, \hepth{9111035}.}
\lref\MOORE{G. Moore, ``Gravitational phase transitions
  and the sine-Gordon model", \hepth{9203061}.}
\lref\DMP{R.Dijkgraaf, G.Moore, and M.R.Plesser,
``The partition function of 2d string theory'',
\np{394}{1993}{356}, \hepth{9208031}.}

\lref\POLCHINSKI{J. Polchinski, ``What is string theory'',
{\it Lectures presented at the 1994 Les Houches Summer School
``Fluctuating Geometries in Statistical Mechanics and Field Theory''},
\hepth{9411028}.}
\lref\KLEBANOV{I. Klebanov, {\it Lectures delivered at the ICTP
Spring School on String Theory and Quantum Gravity},
Trieste, April 1991, \hepth{9108019}.}
\lref\JEVICKI{A. Jevicki, ``Developments in 2D string theory'',
\hepth/9309115.}
\lref\HSU{E. Hsu and D. Kutasov, ``The Gravitational Sine-Gordon
Model'',
\np{396}{1993}{693}, \hepth{9212023}.}
\lref\MSS{G. Moore, N. Seiberg, M. Staudacher }
\lref\MuchiImbimbo{C. Imbimbo and S. Mukhi,
``The topological matrix model of c=1 String",
\hepth{9505127}.}

\lref\JM{M. Jimbo and T. Miwa, ``Solitons and Infinite Dimensional
Lie Algebras'', {\it Publ. RIMS, Kyoto Univ.} {\bf 19}, No. 3
(1983) 943.}
\lref\Hir{R. Hirota, Direct Method in Soliton Theory {\it Solitons},
Ed. by R.K. Bullogh and R.J. Caudrey, Springer, 1980.}
\lref\UT{K. Ueno and K. Takasaki, ``Toda Lattice Hierarchy":
in `Group representations and systems of differential equations',
{\it Adv. Stud. Pure Math.} {\bf 4} (1984) 1.}
\lref\Takasak{K. Takasaki,
{\it Adv. Stud. Pure Math.} {\bf 4} (1984) 139.}
\lref\MukhiImbimbo{ C. Imbimbo and S. Mukhi,
``The topological matrix model of c=1 String",
\np{449}{1995}{553}, \hepth{9505127}.}
\lref\NTT{ T.Nakatsu, K.Takasaki, and S.Tsujimaru,
``Quantum and classical aspects of deformed $c=1$ strings'',
\np{443}{1995}{155}, \hepth{9501038}.}
\lref\Takebe{T. Takebe, ``Toda lattice hierarchy and conservation
laws'',
\cmp{129}{1990}{129}.}
\lref\EK{T. Eguchi and H. Kanno,
``Toda lattice hierarchy and the topological description of the
$c=1$ string theory'', \pl{331}{1994}{330}, hep-th/9404056.}
\lref\Krichever{I. Krichever, {\it Func. Anal. i ego pril.},
{\bf 22:3} (1988) 37  (English translation:
{\it Funct. Anal. Appl.} {\bf 22} (1989) 200);
``The  $\tau$-function of the
universal Witham hierarchy, matrix models and topological field
theories'', {\it Comm. Pure Appl. Math.} {\bf 47} (1992),
\hepth{9205110}.}
\lref\TakTakb{K.~Takasaki and T.~Takebe,
``Quasi-classical limit of Toda hierarchy and W-infinity symmetries'',
\lmp{28}{93}{165}, \hepth{9301070}.}
\lref\orlshu{A. Orlov and E. Shulman, \lmp{12}{1986}{171}.}
\lref\Nakatsu{T. Nakatsu, ``On the string equation at $c=1$'',
\mpl{A9}{1994}{3313}, \hepth{9407096}.}
\lref\TakSE{K. Takasaki,
``Toda lattice hierarchy and generalized string equations'',
\cmp{181}{1996}{131}, \hepth{9506089}.}
\lref\TakTak{K. Takasaki and T. Takebe,
``Integrable Hierarchies and Dispersionless Limit'',
\rmatp{7}{1995}{743}, \hepth{9405096}.}
\lref\Moad{M. Adler and P. van Moerbeke, ``String-Orthogonal
Polynomials, String equations, and 2-Toda symmetries'', {\it Commun.
Pure Appl. Math.} {\bf 50} (1997)
241, \hepth{9706182}.}
\lref\BR{A. Boyarsky and O. Ruchayskiy,
``Integrability in SFT and new representation of KP tau-function'',
\hepth{0211010}. }

\lref\kkvwz{ I. Kostov, I. Krichever, M. Mineev-Veinstein,
P. Wiegmann, and  A. Zabrodin, ``$\tau$-function
for analytic curves", 
{\it Random matrices and their applications, MSRI publications} 
{\bf 40} (2001) 285 (Cambridge University Press),
\hepth{0005259}.}
\lref\Zabrodin{ A. Zabrodin, ``Dispersionless limit of Hirota
equations in some problems of complex analysis'',
\tmatp{129}{2001}{1511}; \tmatp{129}{2001}{239}, {\tt
math.CV/0104169}.}
\lref\bmrwz{ A. Boyarsky, A. Marshakov,  O. Ruchhayskiy,
P. Wiegmann, and  A. Zabrodin, ``On associativity equations in
dispersionless integrable hierarchies",
\pl{515}{2001}{483}, \hepth{0105260}.}
\lref\wz{P. Wiegmann and  A. Zabrodin, ``Conformal maps and
dispersionless integrable hierarchies", \cmp{213}{2000}{523},
\hepth{9909147}.}
\lref\WZ{P. Wiegmann and  A. Zabrodin, ``Large scale correlations
in normal and general non-Hermitian matrix ensembles",
\hepth{0210159}.}
\lref\FERTIGone{ Fertig-1 }
\lref\FERTIGtwo{ Fertig-2 }
\lref\WIEGAGAM{O. Agam, E. Bettelheim, P. Wiegmann, and A. Zabrodin,
``Viscous fingering and a shape of an electronic droplet in
the Quantum Hall regime'',  
\prl{88}{2002}{236802}, {\tt cond-mat/0111333}.}

\lref\WITTENGR{E. Witten, ``Ground Ring of two dimensional string
theory'',
\np{373}{1992}{187}, \hepth{9108004}.}
\lref\AKKII{ S. Yu. Alexandrov, V. A. Kazakov, I. K. Kostov,
work in progress. }

\lref\MUKHIVAFA{S. Mukhi and C. Vafa,
``Two dimensional black-hole as a topological coset model of
c=1 string theory'', \np{407}{1993}{667}, \hepth{9301083}.}
\lref\GOSHALVAFA{D. Ghoshal and C. Vafa,
``c=1 String as the Topological Theory of the Conifold'',
\np{453}{1995}{121}, \hepth{9506122}.}
\lref\KZ{V. A. Kazakov and A. Tseytlin,
``On free energy of 2-d black hole in bosonic string theory'',
\jhep{0106}{2001{}021}, \hepth{0104138}.}
\lref\TECHNER{J. Teschner,
``The deformed two-dimensional black hole'',
\pl{458}{1999}{257}, \hepth{9902189}. }
\lref\JAP{T. Fukuda and K. Hosomichi,
``Three-point Functions in Sine-Liouville Theory'',
\jhep{0109}{2001}{003}, \hepth{0105217}. }

\lref\DVa{R. Dijkgraaf and C. Vafa,
``Matrix Models, Topological Strings, and Supersymmetric Gauge
Theories'',
\np{644}{2002}{3}, \hepth{0206255}. }
\lref\DVb{R. Dijkgraaf and C. Vafa,
``On Geometry and Matrix Models'',
\np{644}{2002}{21}, \hepth{0207106}. }
\lref\DVc{R. Dijkgraaf and C. Vafa,
``A Perturbative Window into Non-Perturbative Physics'',
\hepth{0208048}. }
\lref\DVd{R. Dijkgraaf and C. Vafa,
``$\CN=1$ Supersymmetry, Deconstruction and Bosonic Gauge Theories'',
\hepth{0302011}. }

\lref\IKonf{I. K. Kostov,
``Conformal Field Theory Techniques in Random Matrix models'',
\hepth{9907060}.}



\Title{
\rightline{
\vbox{\baselineskip12pt\hbox
{SPhT-t03/013}\hbox{LPTENS-03/05}\hbox{RUNHETC-2003-04} }}}
{\vbox{\centerline{2D String Theory as  Normal Matrix Model}
\centerline{   }
}}
\vskip -1cm
 \centerline{Sergei Yu. Alexandrov,$^{123}$\footnote{$^{\ast}$}
{alexand@spht.saclay.cea.fr}
Vladimir A.
Kazakov$^{14}$\footnote{$^{\circ}$}{{kazakov@physique.ens.fr}}
and Ivan K. Kostov$^{2}$\footnote{$^{\dagger}$}{Associate member of 
{ \ninepoint \cyr Institut za 
Yadreni Izsledvamiya i Yadrena Energetika},  Bulgarian Acadrmy of 
Sciences,
Sofia, Bulgaria}\footnote{$^\bullet$}{{kostov@spht.saclay.cea.fr}
}}
{ \ninepoint
\centerline{$^1${\it  Laboratoire de Physique Th\'eorique de l'Ecole
Normale Sup\'erieure,\footnote{$^\ast$}{
Unit\'e de Recherche du
Centre National de la Recherche Scientifique et de  l'Ecole Normale
Sup\'erieure et \`a l'Universit\'e de Paris-Sud.} }}
\centerline{{\it  24 rue Lhomond, 75231 Paris CEDEX, France}}
\centerline{$^2${\it Service de Physique Th\'eorique,
CNRS - URA 2306, C.E.A. - Saclay,}}
\centerline{  \it F-91191 Gif-Sur-Yvette CEDEX, France}
\centerline{$^3$ \it V.A.~Fock Department of
Theoretical Physics,  St.~Petersburg
University,}
\centerline{ \it 198904 St.~Petersburg, Russia}
\centerline{$^4$ {\it Dept. of Physics and Astronomy,
Rutgers University, }}
\centerline{\it Piscataway NJ 08855 USA}
}

  \vskip 1cm
\baselineskip8pt{
We show that the $c=1$ bosonic string theory at finite temperature
has two matrix-model realizations
related by a kind of duality transformation.
The first realization is the standard one given by the
compactified matrix quantum mechanics in the inverted oscillator
potential. The second realization,
which we derive here, is given by the normal
matrix model. Both matrix models  exhibit the Toda integrable
structure and are associated with two dual cycles
(a compact and a non-compact one) of
a complex curve with the topology of a sphere with two punctures.
The equivalence of the two matrix models holds for an arbitrary
tachyon perturbation and in all orders in the string coupling
constant.

\baselineskip12pt{
\noindent
}

\Date{}

\baselineskip=14pt plus 2pt minus 2pt

\newsec{ Introduction}

 The $c=1$ string theory\foot{As a good review we recommend
 \KLEBANOV.} has been originally constructed in the early 90's as the
 theory of random surfaces embedded into a one-dimensional spacetime
 \KAZMIG. Since then it became clear that this is only one of the
 realizations of a more universal structure, which reappeared in
 various mathematical and physical problems. Most recently, it was
 used for the description of the 2D black hole
 \refs{\KKK,\AK,\AKTBH,\IK} in the context of the topological string
 theories on Calabi-Yau manifolds with vanishing cycles \GOSHALVAFA\
 and $\CN=1$ SYM theories \DVd.
    
The $c=1$ string theory can be constructed as the collective field
theory for a one-dimensional $N\times N$ hermitian matrix field theory
known also as {\it Matrix Quantum Mechanics} (MQM).  This construction
represents the simplest example of the strings/matrix correspondence.
The collective excitations in the singlet sector of MQM are massless
"tachyons" with various momenta, while the non-singlet sectors contain
also winding modes.
 
The singlet sector of MQM can be reduced to a system of $N$
nonrelativistic fermions in the upside-down gaussian potential. Thus
the elementary excitations of the $c=1$ string can be represented as
collective excitations of free fermions near the Fermi level.  The
tree-level string-theory $S$-matrix can be extracted by considering
the propagation of infinitesimal ``pulses" along the Fermi sea and
their reflection off the ``Liouville wall" \refs{\POLCHINSKI,
\JEVICKI}. The (time-dependent) string backgrounds are associated with
the possible profiles of the Fermi sea \AKK.
  
An important property of the $c=1$ string theory is its integrability.
The latter has been discovered by Dijkgraaf, Moore and Plesser \DMP\
in studying the properties of the tachyon scattering amplitudes.  It was 
demonstrated in \DMP\ that the partition function of the $c=1$
string theory in the case when the allowed momenta form a lattice, as
in the case of the compactified Euclidean theory, is a tau function of
Toda hierarchy \TakTak.  The operators associated with the momentum
modes in the string theory have been interpreted in \DMP\ as Toda
flows. A special case represents the theory compactified at the
self-dual radius $R=1$. It is equivalent to a topological theory that
computes the Euler characteristic of the moduli space of Riemann
surfaces \MUKHIVAFA.  When $R=1$ and only in this case, the partition
function of the string theory has alternative realization as a
Kontsevich-type model \refs{\DMP,\MukhiImbimbo}.

In this paper we will show that there is another realization of the
$c=1$ string theory by the so-called {\it Normal Matrix Model} (NMM). 
This is a complex $N\times N$ matrix model in which the integration
measure is restricted to matrices that commute with their hermitian
conjugates. The normal matrix model has been studied in the recent
years in the context of the laplacian growth problem, the integrable
structure of conformal maps, and the quantum Hall droplets \refs{
\kkvwz, \Zabrodin, \bmrwz, \wz, \WZ, \WIEGAGAM}.  It has been noticed
that both matrix models, the singlet sector of MQM and NMM, have
similar properties. Both models can be reduced to systems of
nonrelativistic fermions and possess Toda integrable structure.  There
is however an essential difference: the normal matrix model describes
a compact droplet of Fermi liquid while the Fermi sea of the MQM is
non-compact.  Furthermore, the perturbations of NMM are introduced by
a matrix potential while these of MQM are introduced by means of
time-dependent asymptotic states.  In this paper we will show that
nevertheless these two models are equivalent.

The exact statement is that the matrix quantum mechanics compactified
at radius $R$ is equivalent to the normal matrix model defined by
the probability distribution function $ \exp[\bh W_{R}¥(Z, \bZ)]$,
where 
\eqn\NMM{W_{R}¥(Z,\bZ)=\tr (Z\bZ)^R -\hb{R-1\over 2}\tr \log
(Z\bZ) - \sum_{k\ge 1} t_k \tr Z^k - \sum_{k\ge 1}t_{-k}\tr{\bZ}^k . 
} 
More precisely, the grand canonical partition function of MQM is
identical to the canonical partition function of NMM at the perturbative
level, by which we mean that the genus expansions of the two free
energies coincide.  This will be proved in two steps.  First, we will
show that the non-perturbed partition functions are equal.  Then we
will use the fact that both partition functions are $\tau$-functions
of Toda lattice hierarchy, which implies that they also coincide in
presence of an arbitrary perturbation.
  
We also give a unified geometrical description of the two models in
terms of a complex curve with the topology of a sphere with two
punctures. This complex curve is analog of the Riemann surfaces
arising in the case of hermitian matrix models. However, for generic
$R$ it is not an algebraic curve. The curve has two dual
non-contractible cycles, which determine the boundaries of the
supports of the eigenvalue distributions for the two models. The
normal matrix model is associated with the compact cycle while the MQM
is associated with the non-compact cycle connecting the two punctures.
We will construct a globally defined one-form whose integrals along
the two cycles give the number of eigenvalues $N$ and the derivative
of the free energy with respect to $N$.

The paper is organized as follows.  In the next section we will remind
the realization of the $c=1$ string theory as a fermionic system with
chiral perturbations worked out in \AKK. We will stress on the
calculation of the free energy, which will be needed to establish the
equivalence with the NMM. In Sect. 3 we construct the NMM having the
same partition function. In Sect. 4 we consider the quasiclassical
limit and give a unified geometrical description of the two models.

\newsec{Matrix quantum mechanics, free fermions and integrability }

\subsec{Eigenfunctions and fermionic scattering}

In absence of winding modes, the 2D string theory is described by the
singlet sector of Matrix Quantum Mechanics in the double scaling limit
with Hamiltonian
 \eqn\MQMH{ {\bf \hat H}_0=\hf \tr(-\hb^{2}{\p^2\over \p X^2}-X^2).  }
The radial part of this Hamiltonian is expressed in terms of the
eigenvalues $x_1,\ldots,x_N$ of the matrix $X$.  The wave functions in
the $SU(N)$-singlet sector are completely antisymmetric and thus
describe a system of $N$ nonrelativistic fermions in the inverse
gaussian potential.  The dynamics of the fermions is governed by the
Hamiltonian
\eqn\oneH{ {\bf \hat H}_0 =\hf \sum_{i=1}^N(\hat p_i^2-\hat x_i^2),
}
where $p_i$ are the momenta conjugated to the fermionic coordinates
$x_i$.

To describe the incoming and outgoing tachyonic states, it is
convenient to introduce the "light-cone" coordinates in the phase
space \eqn\lccor{ \hat\xpm={\hat x\pm \hat p \over \sqrt{2}} }
satisfying the canonical commutation relations
\eqn\ygrekpm{
[ \hat \xp,\hat \xm]=-i\hb .}
In these variables the one-particle Hamiltonian takes the form
\eqn\oneph{ \hat H_0 = -\hf (\hat \xp\hat \xm +\hat \xm\hat \xp).  }
We can work either in $\xp$ or in $\xm$ representation, where the
theory is defined in terms of fermionic fields $\psipm(\xpm)$
respectively.  The solutions with a given energy are
$\psepm(\xpm,t)=e^{- \ibh Et}\psepm(\xpm)$ with 
\eqn\wavef{
\psepm(\xpm)= {1\over\sqrt{2\pi\hb }}\ e^{\mp {i\over 2 \hb } \phi_0}
\xpm^{ \pm{i\over\hb} E-\hf},} 
where the phase factor $\phi_0(E)$ will
be determined below.  These solutions form two complete systems of
$\delta$-function normalized orthonormal states under the condition
that the domain of the definition of the wave functions is a
semi-axis.\foot{In order to completely define the theory, we should
define the interval where the phase-space coordinates $\xpm$ are
allowed to take their values.  One possibility is to allow the eigenvalues
to take any real value.  The whole real axis corresponds to a theory
whose Fermi sea consists of two disconnected components on both sides
of the maximum of the potential.  To avoid the technicalities related
to the tunneling phenomena between the two Fermi seas, we will define
the theory by restricting the eigenvalues to the positive real axis. 
The difference between the two choices is seen only at the
non-perturbative level.}  The left and right representations are
related to each other by a
unitary operator which is the Fourier transform on  the half-line
\eqn\Fourtr{[\hat S \psip](\xm)=  \int_{0}^{\infty}
d\xp \, S(\xm,\xp) \psip(\xp).}
The integration kernel $S(\xm,\xp)$ is either a sine or a cosine
depending on the boundary conditions at the origin. For 
definiteness, let us choose the cosine kernel
\eqn\Scos{ S(\xm,\xp)= \sqrt{2\over \pi\hb} \cos({1\over \hb} 
\xp\xm).}
Then the operator $\hat S$
is diagonal on the eigenfunctions \wavef\ with given energy
\eqn\Smtrxx{[\hat S \psep ](\xm) =
e^{-{i\over\hb}\phi_0}\CR(E) \psem (\xm), }
where
\eqn\rfactor{
\CR(E) = \hb^{iE}\sqrt{2\over \pi}
\cosh\left({\pi\over 2 }
 \left(\bh E- i/2\right)\right)  \Gamma\( \ibh E + 1/2\).
}
Since the operator $\hat S$ is unitary, i.e.
\eqn\unitrtyr{ \overline{\CR(E) }
\CR(E) = \CR(-E) \CR(E)=1,}
it can be absorbed into the function $\phi_0(E)$.  This fixes the
phase of the eigenfunctions as \eqn\fR{ \phi_0(E)=-i\hb \log \CR(E). 
} 
In fact, the operator $\hat S$ appears in our formalism as the fermionic $S$-matrix
describing the scattering off the inverse oscillator potential 
and the factor $\CR(E)$ gives the fermionic reflection coefficient.

Let us introduce the scalar product between left and right states as 
follows
\eqn\scpr{
 \langle\psim|\hat S|\psip\rangle
=\int\!\!\!\!\int_{0}^\infty d\xp d\xm\,
\overline{\psim(\xm)}  S(\xm,\xp)  \psip(\xp).
}
Then the  in- and out-eigenfunctions are orthonormal with respect to 
this
scalar product
\eqn\scprEE{
\langle\psem|\hat S|\peep\rangle= \delta(E-E').}
 It is actually  this relation that defines the phase factor
containing all information about the   scattering.

The ground state of the Hamiltonian \oneph\ can be constructed from
the wave functions \wavef\ and describes the linear dilaton background
of string theory \JEVICKI. The tachyon perturbations can be introduced
by changing the asymptotics of the wave functions at $x_\pm\to \infty$
to
\eqn\asswave{
\Pepm(\xpm)\sim
e^{\mp \ibh R\sum\limits t_{\pm k} \xpm^{k/R}}
\xpm^{\pm \ibh E-\hf}.}
The exact phase contains also a constant mode $\mp {1\over 2\hb}
\phi(E)$ as in \wavef\ and negative powers of $\xpm^{1/R}$ which are
determined by the orthonormality condition
\eqn\orth{
\langle  \Pim{E}|\hat S|  \Pip{E'}\rangle =\delta(E-E').}
The constant mode $\phi(E)$ of the phase of the fermion wave function 
contains all essential information about the perturbed system.

\subsec{Cut-off prescription and  density of states  }

To find the density of states, we introduce a cut-off $\Lambda $ by
confining the phase space to a periodic box \eqn\ctfF{ \xpm +
2\sqrt{\L} \equiv \xpm.} which can be interpreted as putting a
reflecting wall at distance $\sqrt{\L} $.  This means that at the
points $\xp= \sqrt{\L} $ and $\xm= \sqrt{\L} $ the reflected wave
function coincides with the incoming one 
\eqn\bcctff{ [\hat
S\Psi](\ctf) =\Psi(\ctf).  } 
Applied to the wave functions \asswave,
this condition gives an equation for the admissible energies
\eqn\kvcon{e^{\ibh \phi(E)}=e^{ -\ibh V(\Lambda) } \Lambda^{\ibh E},
\qquad V(\Lambda)=R\sum\limits (t_{k}+t_{-k})\Lambda^{k/2R} } 
where we
neglected the negative powers of $\Lambda$.  It is satisfied by a
discrete set of energies $E_n$ defined by the relation \eqn\kvconbis{
E_n\log \Lambda-\phi(E_n)=2\pi\hb n+V(\Lambda),\ \ n\in \Zb.  } Taking
the limit $\Lambda \rightarrow \infty$ we find the density of states
(in units of $\hb$)
\eqn\DEN{
\rho(E)={\log \Lambda\over 2 \pi}-{1\over 2\pi } {d\phi(E) \over d E}.
}

\subsec{The partition function of the compactified 2D string theory}

Knowing the density of states, we can calculate the grand canonical
partition function $\ZM$ of the quantum-mechanical system
compactified at Euclidean time
$\b= 2\pi R$.  If $\mu$ is the chemical potential, then the free
energy $\CF = \hb^{2}\log \ZM$ of the ensemble of free fermions is
given by
\eqn\FRENO{
\CF(\mu,t)=\hb \int_{-\infty}^\infty
d E\, \rho(E)\log\left[1+e^{-\bh\beta(\mu+E)}\right].
}
Integrating by parts and dropping out the $\Lambda$-dependent
non-universal piece, we obtain 
\eqn\FRENOo{ \CF(\mu,t)=-{\beta\over
2\pi }\int_{-\infty}^{\infty} dE\, {\phi(E)\over 1+e^{\bh
\beta(\mu+E)}} .  } 
We close the contour of integration in the upper
half plane and take the integral as a sum of residues of the thermal
factor\foot{We neglect the nonperturbative terms associated with the
cuts of the function $\phi(E)$.}
\eqn\FRENooo{
\CF(\mu,t)=i\hb \sum\limits_{n \ge 0}
\phi\left(i\hb{n+\hf\over R}-\mu\right).
}

We will be interested only in the perturbative expansion in $1/\mu$
and neglect the non-perturbative terms $\sim e^{-{\pi\over \hb} \mu}$,
as well as non-universal terms in the free energy (regular in $\mu$).
Therefore, for the case of zero tachyon couplings we can retain only
the $\Gamma$-function in the reflection factor \rfactor \ and choose
the phase $\phi_0$ as
\eqn\phino{   \phi_0(E)=
-i\hb  \log  \Gamma\(\ibh E + 1/2\).}

It is known \DMP\ that the tachyon scattering data for MQM are 
generated by a  $\tau$-function of Toda lattice hierarchy where the coupling
constants $t_k$ play the role of the Toda times.
The $\tau$-function  is related to the constant
mode $\phi(E)$ by \AKK
\eqn\phitau{e^{{i\over \hb} \phi(-\mu)}= {\taum_{0}\(\mu +i{\hbar\over
2R}\)\over \taum_{0}\(\mu -i{\hbar\over 2R}\)}.  }
Taking into account \FRENooo\ we conclude that the partition function 
of the perturbed MQM  is equal to the $\tau $-function: 
\eqn\Ztauu{\ZM (\mu,t)=\taum_{0}(\mu,t).}
The discrete space parameter $s\in\Z$ along the Toda chain is related
to the chemical potential $\mu$.  More precisely, $s$ corresponds to
an imaginary shift of $\mu$ \refs{\IK,\AKK}:
\eqn\SHIFT{ \taum_{s}(\mu,t) =\taum_{0}\(\mu+i\hb {s\over R},t \).   }
This fact was used to rewrite the Toda equations as difference
equations in $\mu$ rather than in the discrete parameter $s$. 
It will be also used below to prove the equivalence of 2D string theory
to the normal matrix model.

\newsec{  2D string theory as normal matrix model}

In this section we will show that the partition function of 2D string theory
with tachyonic perturbations can be rewritten as a normal matrix
model. It is related to the original matrix quantum mechanics by a
kind of duality transformation. In fact, we will give two slightly
different normal matrix models fulfilling this goal.

Consider the following matrix integral
\eqn\matr{
\ZN_{\hb}(N,t,\alpha)=\int\limits_{[Z,\bZ]=0} d Z d \bZ\,
 \left[\det (Z\bZ)\right]^{ (R-1)/2+{\alpha\over \hb}}
e^{-\bh \tr[  (Z \bZ)^R - V_+(Z)-
V_-(\bZ)]},
}
where  the integral goes over all complex matrices satisfying
$[Z,\bZ]=0$ and the potentials are given by
\eqn\Vbig{V_\pm(z)
= \sum\limits_{k\ge 1} t_{\pm k} z^{k} . }
We made the dependence on the Planck constant explicit because later
we will need to analytically continue in $\hb$.  The integral \matr\
defines a $\tau$-function of Toda hierarchy \WZ. We remind the proof
of this statement in Appendix A, where we also derive the string
equation specifying the unique solution of Toda equations.  This
string equation coincides (up to change $\hb\to i\hb$) with that of
the hierarchy describing the 2D string theory \refs{\IK,\AKK}. 
Therefore, if we identify correctly the parameters of the two models, the
$\tau$-functions should also be the same.  Namely, we should relate
the chemical potential $\mu$ of 2D string theory to the size of
matrices $N$ and the parameter $\alpha$ of the normal matrix model. 
There are two possibilities to make such identification.

\subsec{Model I}

The first possibility is realized taking a large $N$ limit of the
matrix integral \matr.  Namely, we will prove that the full perturbed
partition function of MQM is given by the large $N$ limit of the
partition function \matr\ with $\a = R\mu-\hb N$ and a subsequent
analytical continuation $\hb\to i\hb$
\eqn\PARTFU{ \ZM _{\hb}(\mu,t)= \lim\limits_{N\to \infty}
\ZN _{i\hb}(N,t,R\mu-i\hb N).}
The necessity to change the Planck constant by the imaginary one
follows from the comparison of the string equations of two models as
was discussed above.  More generally, the $\tau$-function \SHIFT\ 
for  arbitrary $s$ is obtained from the partition function \matr\ in the
following way 
\eqn\PARtau{ \taum_{s,\hb}(\mu,t)= \lim\limits_{N\to \infty}
\ZN_{i\hb}(N+s,t,R\mu-i\hb N).}
Let us stress that in spite of the large $N$ limit taken here, we
obtain as a result the full (and not only dispersionless) partition
function of the $c=1$ string theory.

To prove this statement we will use the fact that both partition
functions are $\tau$-functions of the Toda lattice hierarchy with
times $t_{\pm k}$,~$k=1,2,\ldots$.  Since the unperturbed
$\tau$-function provides the necessary boundary conditions for the
unique solution of Toda equations, it is sufficient to show that the
unperturbed partition functions of the two models coincide and the
$s$-parameters of two $\tau$-functions are identical.

The integral \matr\ reduces to the product of
the normalization coefficients of the orthogonal polynomials, eq. 
(A.5). From the definition of the scalar product (A.4) it follows
that when all $t_{\pm k}=0$ the orthogonal polynomials are simple monomials and
the normalization factors $h_n$ are given by
\eqn\hn{
h_n(0,\alpha)={1\over 2\pi i}
 \int_{\Cb} d^2 z \,
e^{-\bh(z\bz)^R}  (z\bz)^{(R-1)/2+{\alpha\over \hb}+n}.
}
Up to constant factors and powers of $\hb$, we find
\eqn\hnRfac{
h_n (0,\alpha)\sim
\Gamma\left( {\alpha\over \hb R} +{n+\hf\over R}+\hf\right).}
We see that the result coincides with the reflection coefficient
given in \phino.
Therefore, taking the analytical continuation of parameters as
in \PARTFU, we have
\eqn\provefs{\eqalign{
\lim\limits_{N\to \infty} \hb^2 \log \ZN_{i\hb}(N,0,R\mu-i\hb N)&=
\lim\limits_{N\to \infty} i\hb \sum\limits_{n=0}^{N-1}
\phi_0\left(i\hb  {N-n-\hf \over R}-\mu\right)
\cr
&=i\hb \sum\limits_{n=0}^{\infty}
\phi_0\left( i\hb  {n+\hf \over R}-\mu\right).
}}
This coincides with the unperturbed free energy $\CF(\mu,0)$ from
\FRENooo, what proves \PARTFU\ for all $t_{\pm k}=0$.

Thus, it remains to show that the $s$-parameter  of the $\tau$-function
describing the normal matrix model is associated with $\mu$.
From (A.17) we see that it coincides with $N$. Hence
it trivially follows from  \PARtau\ that
\eqn\taumu{\taum_{s,\hb}(\mu,t)=
\lim\limits_{N\to \infty}\ZN_{i\hb}\(N,t,R\( \mu+i\hb {s\over R}\)-i\hb N\)=
\taum_{0,\hb}\(\mu+i\hb {s\over R},t\), }
what means that the $\tau$-function defined in \PARtau\ possesses
the characteristic property \SHIFT.

\subsec{Model II}

In fact, one can even simplify the representation \PARTFU\ what
will provide us with the second realization of 2D string theory
in terms of NMM.
Let us consider the matrix model \matr\ for $\alpha=0$.
We claim that the partition function $Z_{i\hb}(N,t,0 )$ in the
canonical ensemble, analytically continued to imaginary Planck constant and
\eqn\numb{
N=-\ibh R\mu,
}
coincides with the partition function of 2D string theory in the grand canonical
ensemble:\foot{From now on  we will omit the last argument of $\ZN$
corresponding to the vanishing parameter~$\alpha$.}
\eqn\identZZ{ \ZM_{\hb}(\mu,t)=\ZN_{i\hb}(-\ibh R\mu,t).}

Indeed, as for the first model, they are both given by
$\tau$-functions of Toda hierarchy.  After the identification \numb,
the $s$-parameters  of these $\tau$-functions are identical due to (A.17) and
\SHIFT. Therefore, it only remains to show that the integral
$Z_{i\hb}(N,0)$ without potential is equal to the unperturbed 
partition
function \FRENooo.  In this case the method of orthogonal polynomials
gives
\eqn\matrwww{
\ZN_{i\hb}(N,0)=\prod\limits_{n=0}^{N-1}\CR\left(-i\hb {n+\hf\over 
R}\right), }
where the $\CR$ factors are the same as in MQM, as was shown in 
\hnRfac.
Then we write
\eqn\zzzzz{
 \ZN_{i\hb}(N,0)=\Xi (0)/\Xi (N),\qquad {\rm where} \quad
\Xi (N)=\prod\limits_{n=N}^{\infty}\CR\left(-i\hb (n+\hf)/R\right). }
$\Xi (0)$ is a constant and can be neglected, whereas $\Xi (N)$ can be
rewritten as
\eqn\zzzz{ \Xi
(N)=\prod\limits_{n=0}^{\infty}\CR\left(-i\hb N/R-i\hb
(n+\hf)/R\right).  } 
Taking into account the unitarity relation
\unitrtyr\ and substituting $N$ from \numb, we obtain 
\eqn\ZZZ{
\ZN_{i\hb}(-\ibh R\mu,0)\sim \Xi ^{-1}(-\ibh
R\mu)=\prod\limits_{n=0}^{\infty}\CR \left(\mu+i\hb (n+\hf)/R\right). 
} 
Thus $\hb^2\log \ZN_{i\hb}(-\ibh R\mu,0)$ coincides with the
free energy \FRENooo\ with  all $t_k=0$.  Note that the difference in
the sign of $\mu$ does not matter since the free energy is an even
function of $\mu$ (up non-universal terms).\foot{Moreover, this sign
can be correctly reproduced from \FRENOo\ if closing the integration
contour in the lower half plane.} Since the two partition functions 
are both solutions of the Toda hierarchy, the fact that they coincide 
at $t_{k}=0$ implies that they coincide for arbitrary perturbation.

We conclude  that the grand canonical partition function
of the $c=1$ string theory  equals the 
canonical partition function of the normal matrix model \matr\ for
$\alpha=0$ and $N=-\ibh R\mu$.   It is also  clear that
we can  identify the operators of tachyons in two models
\eqn\idop{ \Tr \Xp^{n/R} \leftrightarrow \Tr Z^{n}, \qquad \Tr
\Xm^{n/R} \leftrightarrow \Tr (\bZ)^{n}.  }

\newsec{ Geometrical meaning of the
duality between the two models}

The quasiclassical limit of most of the solvable matrix models has a
nice geometrical interpretation.  Namely, the free energy in this
limit can be parameterized in terms of the periods of a holomorphic
1-form around the cycles of an analytical curve \refs{\DAVID,\DavidEy,
\IKonf, \KM}.  Recently this geometrical picture appeared also in the
context of the topological strings on singular Calabi-Yau manifolds
and supersymmetric gauge theories \refs{\DVa, \DVb, \DVc}.

In this section we will show that both MQM and NMM have in the
quasiclassical limit   a similar geometrical interpretation in terms of
a one-dimensional complex curve, which is topologically a sphere with
two punctures.  Here by complex curve we understand a complex manifold
with punctures and given behavior of the functions on this manifold at
the punctures.  Each of the two matrix models corresponds to a
particular real section of this curve coinciding with one of its two
non-contractible cycles, and the duality between them is realized by
the exchange of the cycles of the curve.

\subsec{The dispersionless limit}

The quasiclassical limit $\hb\to 0$ corresponds to the dispersionless
limit of the Toda hierarchy where it has a description in terms of a
classical dynamical system.  The Lax operators are considered as
$c$-functions of the two canonically conjugated coordinates $\mu$ and
$\o$, where $\o$ is the quasiclassical limit of the shift operator
$\hat \omega = e^{\p/\p s}$.  Moreover, the two Lax operators can be
expressed as series in $\o$.  The solutions of the dispersionless
hierarchy correspond to canonical transformations in the phase space
of the dynamical system.  In the case of MQM this is the
transformation from the coordinates $\mu$ and $\log\o$ to $\xp$
and $\xm$. 

The particular solution of the Toda hierarchy that appears in our
problem also satisfies the dispersionless string equation.  In the
case of MQM this is nothing but the equation of the profile of the
Fermi sea in the phase space \AKK
\eqn\TODApr{ \xp\xm=\sum _{k\ge 1} k t_{\pm k} \xpm^{ k/R} +\mu +
\sum_{k\ge 1} v_{\pm k} \xpm^{-k/R}.}
To get the string equation for  NMM it is enough to make the substitution following \idop
\eqn\idvar{ \xp \leftrightarrow z^R,
\qquad \xm \leftrightarrow \bz^R }
and $\mu=\hb N/R$ as explained in the previous section.  (We took into
account that $\hb$ from \numb\ should be replaced here by $-i\hb$ so
that the factor $i$ is canceled.)  Then the string equation describes
the contour $\gamma$ bounding the region $\CD$ in the complex
$z$-plane filled by the eigenvalues of the normal matrix.\foot{Here we
consider the case when the eigenvalues are distributed in a simply
connected domain.}

\subsec{NMM in terms of electrostatic potential}

In this section we will introduce a function of the spectral variables
$z$ and $\bz$, which plays a central role in the NMM integrable
structure.  This function has several interpretations in the
quasiclassical limit.  First, it can be viewed as the generating
function for the canonical transformation mentioned above, which maps
variables $\mu$ and $\log \o$ to the variables $z$ and $\bz$.  Second,
it gives the phase of the fermion wave function at the Fermi level
after the identification \idvar.

There is also a third, electrostatic interpretation, which is
geometrically the most explicit and which we will follow in this
section.\foot{See also \BR\ for another interesting interpretation
which appears useful in string field theory.}  
According to this interpretation, the eigenvalues
distributed in the domain $\CD$ can be considered to form a charged
liquid with the density
\eqn\dens{ \rho(z,\bz)={1\over \pi}
\p_{z}\p_{\bz} W_{R}(z,\bz)={R^2\over \pi}(z\bz)^{R-1}.  }
Let $\vp (z,\bar z)$ be the potential of the charged eigenvalue
liquid, which is a harmonic function outside the domain $\CD$ and it
is a solution of the Laplace equation inside the domain 
\eqn\phiout{
\vp (z,\bar z)=\cases{ \vp (z )+\bar\vp ( \bar z),& $z\not\in \CD$,
\cr (z\bz)^R, & $z\in \CD$.} } 
To fix completely the potential, we should also impose the asymptotics
of the potential at infinity. This asymptotics is determined by the
coupling constants $t_{\pm n}, n=1,2,...$ and can be considered as the
result of placing a dipole, quadrupole etc. charges at infinity.

The solution of this electrostatic problem is obtained as follows.
The  continuity of the potential $\vp(z,  \bar z)$ and its first derivatives 
leads to the following  conditions to be satisfied on the boundary $\g=\p \CD$
 \eqn\dirichl{
\vp (z)+ \bar\vp (\bz)=(z\bz)^R ,} 
\eqn\streq{ z\p_z\vp (z)= \bar z \p_{\bz} \bar\vp
(\bz)=R z^R\bar z^R.  } 
Each of two equations \streq\ can be interpreted as an equation for the
contour $\gamma$. Since we obtain two equations for one curve,
they should be compatible. This imposes a
restriction on the holomorphic functions $\vp(z)$ and $\bar \vp(\bz)$.
The solutions for these chiral parts of the potential can
be found comparing \streq\ with the string equation \TODApr.  In this
way we have
\eqn\vppZ{\eqalign{
\vp(z)&=  \hb N \log z + \hf \phi  +R \sum\limits_{k\ge 1}t_k z^k-R
\sum\limits_{k\ge 1}{1\over k}v_{k} z^{-k}, \cr
\bar \vp(\bz)&= \hb N \log \bz + \hf \phi  +
R\sum\limits_{k\ge 1}t_{-k} \bz^k-
R\sum\limits_{k\ge 1}{1\over k}v_{-k} \bz^{-k} .
}}
The zero mode $\phi$ is fixed by the condition \dirichl. However, it
is easier to use another interpretation. As we mentioned above, the
holomorphic functions $\vp(z)$ and $\bar \vp(\bz)$ coincide with 
the phases of the fermionic wave functions and
in particular their constant modes are the same.  Then, taking the limit
$\hb\to 0$ of \phitau, in the variables of NMM we find
\eqn\zeromod{
\phi =-\hb {\p \over \p N}\log \ZN.
}

\subsec{The complex curve}

 Instead of considering the function $\vp (z,\bz)$ harmonic  in
the exterior domain $\bar \CD= {\bf C}/\CD$, we can introduce
one holomorphic function  $\vvp$ defined in the double cover of $\bar 
\CD$.
The doubly covered domain $\bar \CD$ is topologically equivalent
to a two-sphere with two punctures (at $z=\infty $ and $ 
\bz=\infty$),
which we identify with the north and the south poles.
It can be covered by two coordinate patches associated with the north
and south hemispheres and parameterized by $z$ and $\bz$.
They induce a natural complex structure so that the sphere can
be viewed as a complex manifold.
The patches overlap in a ring containing the contour $\gamma$
and the transition function between them $\bz(z)$ (or $z(\bz)$)
is defined through the string equation \streq.

The string equation can also be considered as a defining equation
for the manifold. We start with the complex plane $\Cb^2$
with flat coordinates $z$ and $\bz$. Then the manifold is defined
as solution of equation \streq, thus providing its natural embedding 
in $\Cb^2$. 

To completely define the complex curve, we should also specify the
singular behavior at the punctures of the analytic functions on this
curve. We fix it to be the same as the one of the holomorphic parts
of the potential \vppZ .
 
We define  on the complex curve the following  holomorphic field 
\eqn\chirf{
\vvp  \mathop{=}\limits^{\rm def}
\cases{\vvp_+(z)= \vp (z)-\hf (z\bz(z))^R
& in the north hemisphere, \cr
\vvp_-(\bz) = -\bar\vp (\bz)+\hf (z(\bz)\bz)^R
& in the south hemisphere.}
} 
Its analyticity  follows from equation \dirichl, which now holds
on the entire curve, since $z$ and $\bz$ are no more considered
as conjugated to each other. 

The field $\vvp$ gives rise to a closed (but not exact) holomorphic
1-form $d\vvp$, globally defined on the complex curve. Actually, this
is the unique globally defined 1-form with given singular behavior at
the two punctures.  Therefore, we can characterize the curve by
periods of this 1-form around two conjugated cycles $A$ and $B$ defined
as follows. The cycle $A$ goes along the ring where both
parameterizations overlap and is homotopic to the contour $\gamma$.
The cycle $B$ is a path going from the puncture at $z=\infty$ to the
puncture at $\bz=\infty$.

The integral around the cycle $A$ is easy to calculate
using \vppZ\ and \streq\ by picking up the pole
\eqn\cyclA{
{1\over 2\pi i} \oint_A d\vvp=
{1\over 2\pi i} \oint_\g d\vvp_+ (z)=
{1\over 2\pi i} \oint_{\g^{-1}} d\vvp_- (\bz)
=\hb N,}
where we indicated that in $\bz$-coordinates the contour should be 
reversed.

To find the integral along the non-compact cycle $B$, one should 
introduce
a regularization by cutting it at $z=\sqrt{\L}$ and $\bz=\sqrt{\L}$.
Then we obtain
\eqn\cyclBi{
 \int_B d\vvp =
\int^{z_0}_{\sqrt{\L}} d\vvp_+ (z)
+\int_{\bz(z_0)}^{\sqrt{\L}} d\vvp_- (\bz)
= \vvp_-(\sqrt{\L})-\vvp_+(\sqrt{\L}),
}
where $z_0$ is any point on the cycle and we used \dirichl.
Taking into account the definition \chirf\ and the explicit solution
\vppZ, we see that the result contains the part vanishing at $\L\to 
\infty$,
the diverging part which does not depend on $N$
and, as non-universal, can be neglected, and the contribution of
the zero mode $\phi $. Since $\phi $ enters $\vvp_+$
and $\vvp_-$ with different signs, strictly speaking, it is not a zero
mode for $\vvp$. As a result, it does not disappear from the 
integral, but is doubled. Finally, using  \zeromod\ we get
\eqn\cyclB{ \int_B d\vvp=\hb {\p \over \p N}\log \ZN. }
In Appendix B we derive this formula using the eigenvalue transfer
procedure similar to one used in \DVc, or in \KM\ in a similar case of
the two matrix model equivalent to our $R=1$ NMM.  Thus, the free
energy of NMM can be reproduced from the monodromy of the holomorphic
1-form around the non-compact cycle on the punctured sphere.

Since the string equations coincide, it is clear that the solution of
MQM is described by the same analytical curve and holomorphic 1-form.
To write the period integrals in terms of MQM variables, it is enough
to change the coordinates according to \idvar\ and $\hb N$ = $R\mu$ in
all formulas. In this way we find\foot{To get the second integral from
\cyclB, one should take into account that also $\hb\to i\hb$.}
\eqn\intMQM{
{1\over 2\pi i} \oint_A d\vvp=R\mu, \qquad
\int_B d\vvp= -{1\over R}{\p \CF\over \p \mu}.
}
The last integral can be also obtained   from the integral over the 
Fermi sea.
Indeed, we can choose a point on the contour of the Fermi sea and
split  the integral, which actually calculates the area of the sea,
into three parts
\eqn\intsea{
{1\over R}{\p \CF\over \p \mu}=-\int_{\rm F.s.}d\xp d\xm=
\int_{x_0}^{\sqrt{\L}}\xm(\xp)d\xp+ \int_{\xm(x_0)}^{\sqrt{\L}}
\xp(\xm)d\xm +x_0\xm(x_0).
}
Since $\p_+ \vp =\xm(\xp)$ and $\p_- \bar\vp=\xp(\xm)$ (see \streq),
the last expression is equal  to the integral \cyclBi\ of $d\vvp$
around the (reversed) cycle $B$.
This derivation gives an independent check that  the free energies 
of NMM and MQM do coincide.

\subsec{Duality}

We found that the solutions of both models are described by the same
complex curve.  The curve is characterized by a pair of conjugated
cycles: the compact cycle $A$ encircling one of the punctures and the
non-compact cycle $B$ connecting the two punctures.  The parameter of
the free energy $\mu$ or $N$ and the derivative of the free energy
itself are given by the integrals of the unique globally defined
holomorphic 1-form along the $A$ and $B$ cycles, correspondingly.

\fig{Symbolic representation of the complex curve and the two real 
sections along the cycles $A$ and $B$. The filled regions symbolize 
the Fermi seas of the two matrix models.}{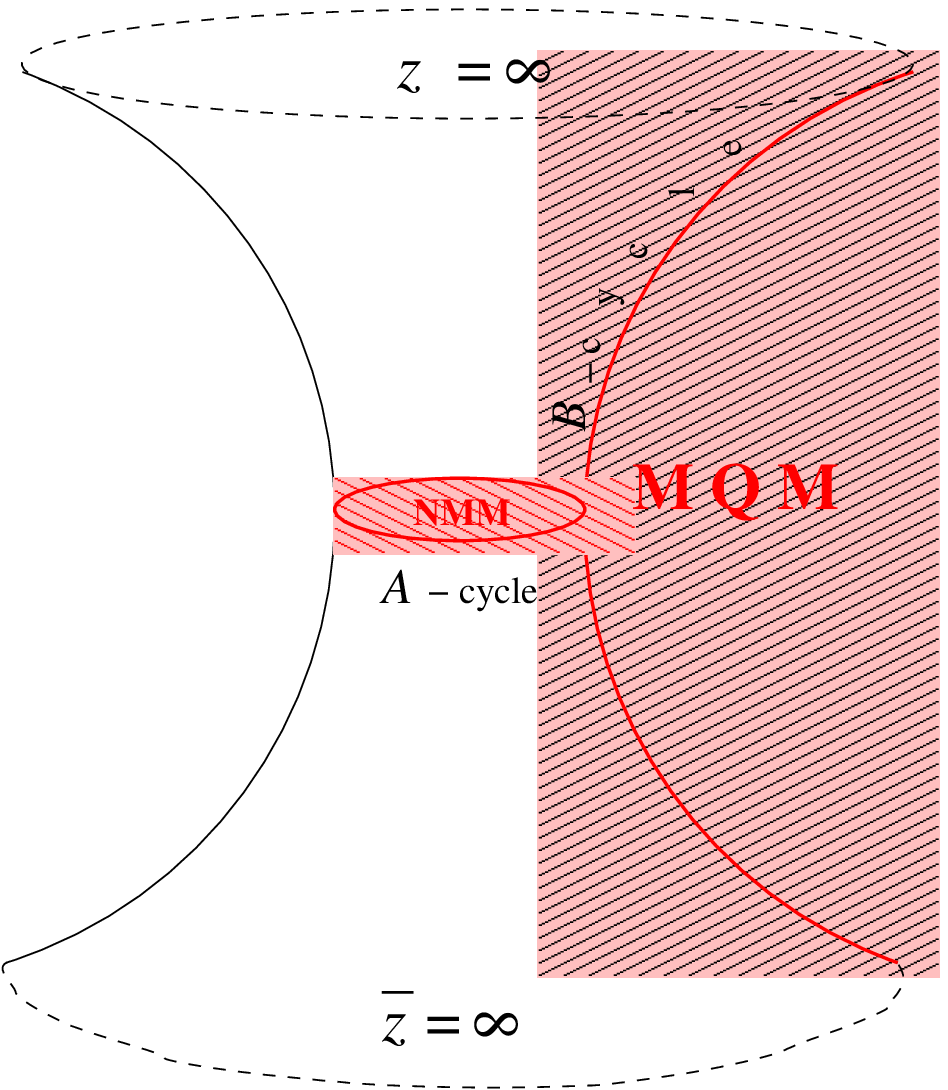}{160pt}

Now one can ask: what is the difference between the two models?  Is it
seen at the level of the curve?  In fact, both models, NMM and MQM can
be associated with two different real sections of the complex curve. 
Indeed, in NMM the variables are conjugated to each other, whereas in
MQM they are real.  Therefore, let us take the interpretation where
the curve is embedded into $\Cb^2$ and consider its intersection with
two planes.  The first plane is defined by the condition $z^*=\bz$ and
the second one is given by $z^*=z$, $\bz^*=\bz$.  In the former case
we get the cycle $A$, whereas in the latter case the intersection
coincides with the cycle $B$ (see fig.  1).  One can think about the
planes as the place where the eigenvalues of NMM and MQM,
correspondingly, live (with the density given by \dens).  
Then the intersections describe the contours
of the regions filled by the eigenvalues.  We see that for NMM it is
given by the compact cycle $A$, and for MQM the Fermi sea is bounded
by the cycle $B$ of the curve.

Therefore, the duality between the two models can be interpreted
as the duality exchanging the cycles of the complex curve
describing the solution. Under this exchange the real variables of
one model go to the complex conjugated variables of another model,
and the grand canonical free energy is replaced by the canonical one.

This duality can be seen even more explicitly if one rewrites the
relations \intMQM\ in terms of the canonical free energy of MQM
defined as $F=\CF+\hb R\mu M$, where $M=-{1\over \hb
R}{\p\CF\over\p\mu}$ is the number of eigenvalues.  Then they take the
following form 
\eqn\intMQMF{ {1\over 2\pi i} \oint_A d\vvp=\bh {\p
F\over \p M}, \qquad \int_B d\vvp= \hb M. } 
They have exactly the same
form as \cyclA\ and \cyclB\ provided we change the cycles.  In
particular, the numbers of eigenvalues are given in the two models as
integrals around the dual cycles.

\newsec{ Conclusions and problems}

We have shown that the the tachyonic sector of the compactified $c=1$
string theory is described in terms of a normal matrix model.  This
matrix model is related by duality to the traditional representation
in terms of Matrix Quantum Mechanics.  More precisely, the canonical
partition function of NMM is equal to the grand canonical partition
function of MQM, provided we identify the number of eigenvalues $N$ of
NMM with the chemical potential $\mu$ in MQM. This holds for any
compactification radius.  Moreover, the duality between the two matrix
models has a nice geometrical interpretation.  In the quasiclassical
limit the solutions of the two models can be described in terms of a
complex curve with the topology of a sphere with two punctures.  The
Matrix Quantum Mechanics and the Normal Matrix Model are associated
with two real sections of this curve which are given by the two
non-contractible cycles on the punctured sphere.  The duality acts by
exchanging the two cycles.

The duality we have observed relates a fermionic problem with
non-compact Fermi sea and continuous spectrum to a problem with
compact Fermi sea and discrete spectrum.  The solutions of the two
problems are obtained from each other by analytic continuation.  Thus,
the reason for the duality can be seen in the common analytic
structure of both problems.  We also emphasize that the duality
relates the partition functions in the canonical and the grand
canonical ensembles, which are related by Legendre transform.  This is
natural regarding that it exchanges the cycles of the complex curve.

Let us mention two among the many unsolved problems related to the
duality between these two models.  First, it would be interesting to
generalize the above analysis to the case when the eigenvalues of NMM
form two or more disconnected droplets.  For $R=1$ this is the analog
of the multicut solutions of the two-matrix model, considered recently
in \KM. In terms of MQM this situation would mean the appearance of
new, compact components of the Fermi sea.  The corresponding complex
curve has the topology of a sphere with two punctures and a number of
handles.

Second, we would like to generalize the correspondence between the two
matrix models in such a way that it incorporates also the winding
modes.  For this we should understand how to introduce the winding
modes in the Normal Matrix Model.  Whereas in MQM they appear when we
relax the projection to the singlet sector, in NMM this could happen
when we relax the normality condition $[Z,\bar Z]=0$.

Up to now there were two suggestions how to include both the tachyon
and winding modes within a single matrix model.  In \AKK, a 3-matrix
model was proposed with interacting two hermitian matrices and one
unitary matrix.  This model can be seen as a particular reduction of
Euclidean compactified MQM. The model correctly describes the cases of
only tachyon or only winding perturbations, though its validity in the
general case is still to be proven.  For the particular case of the
self-dual radius and the multiples of it, a 4-matrix model was
recently proposed in \DVd .  It is based on the old observation of
\WITTENGR\ about the geometry of the ground ring of $c=1$ string
theory.  It would be interesting to find the relation of this model to
our approach, at least in the above-mentioned particular cases.  The
general understanding of this important problem is still missing.

\bigskip \noindent{\bf Acknowledgements:}  We  are grateful to M. Aganagic,
R. Dijkgraaf, S. Gukov, M. Mari\~no, C. Vafa and P. Wiegmann for
valuable discussions.  We thank for the hospitality Jefferson Physical
Laboratory of Harvard University, where a part of this work has been
done. The work of S.A. and V.K. was partially supported by European
Union under the RTN contracts HPRN-CT-2000-00122 and -00131.  The work
of S.A. and I.K was supported in part by European networks EUROGRID
HPRN-CT-1999-00161 and EUCLID HPRN-CT-2002-00325.


\appendix{A}{  Toda description of the perturbed system }

Here we show that  the partition
function \matr\  is a  $\tau$-function of the Toda lattice
hierarchy (for details see \Moad).
We will do it following the standard method of orthogonal polynomials. 
First, we write the integral \matr\ as integral over the eigenvalues
$z_1, ..., z_N$ 
\eqn\PARTFE{ \ZN_{\hb}(N,t,\alpha)= {1\over N!} \int
\prod_{k=1}^N w_\alpha(z_k,\bz_k) \Delta(z)\Delta(\bz) }
with measure
\eqn\DEFdw{
w_\alpha(z,\bz)= {d^2 z\over 2\pi i} \,
e^{-\bh\left[(z\bz)^R-V_+(z)- V_-(\bz)\right]}
(z\bz)^{(R-1)/2+{\alpha\over \hb}}.
}
Then we introduce a set of bi-orthogonal polynomials
\eqn\PERTF{
\Phip^n(z)=
{1\over n!\sqrt{h_n}} \int_{\Cb}
 \prod_{k=1}^n {w_\alpha(z_k,\bz_k)\over h_{k-1}(t,\alpha)} \,
\Delta(z)\Delta(\bz) {\prod_{k=1}^n (z-z_k) } ,
}
and similarly for $\Phim^{n}(\bz)$. Here
  $\Delta(z)=\prod\limits_{k<l}(z_k-z_l)$
is the Vandermond determinant
and normalization factors $h_k$ are
 determined by the orthonormality condition
\eqn\ORTN{
\langle \Phim^{n}|\Phip^{m}\rangle_{_w} \equiv \int_{\Cb}
w_\alpha(z,\bz)\, \Phim^{n}(\bz)\Phip^{m}(z)
= \d_{n-m}. }
 Then the  partition function \PARTFE\   reduces to the
 product of the normalization factors
\eqn\FhNP{
\ZN_{\hb}(N,t,\alpha)=   \prod\limits_{n=0}^{N-1} h_n(t,\alpha).
}

The operators of multiplication by $z$ and $\bz$
are represented in the basis of orthogonal polynomials by the
infinite matrices
\eqn\matrLax{
z\Phip^{n}(z)=\sum\limits_{m} L_{nm}\Phip^{m}(z), \qquad
\bz\Phim^{n}(\bz)=\sum\limits_{m} \Phim^{m}(\bz)\bL_{mn}.
}
with
\eqn\propLax{\eqalign{
& L_{n,n+1}=\bL_{n+1,n}=\sqrt{h_n/h_{n+1}}, \cr
&\ L_{nm}=\bL_{mn}=0, \quad m>n+1.
}}
Differentiation of the orthogonality relation \ORTN\ with respect to
coupling constants $t_k$ gives
\eqn\relatP{\eqalign{
& \hb{\p \Phip^{n}(z) \over \p t_k}=
-\sum\limits_{m=0}^{n-1}(L^k)_{nm}\Phip^{m}(z)
-\hf  (L^k)_{nn}\Phip^{n}(z),
\cr
& \hb {\p \Phip^{n}(z) \over \p t_{-k}}=
-\sum\limits_{m=0}^{n-1}(\bL^k)_{nm}\Phip^{m}(z)
-\hf  (L^k)_{nn}\Phip^{n}(z)
}}
and similarly for $\Phim^n(\bz)$.
Let us define now a wave function which is a vector $\Psi=\{\Psi_n\}$
with elements
\eqn\elwf{
\Psi_n(t;z)=\Phip^n(z)e^{\bh V_+(z)}.
}
Then the equations \relatP\ lead to the following eigenvalue problem
\eqn\eigpsi{
z\Psi=L\Psi,
\qquad
\hb {\p \Psi \over \p t_k} = H_k\Psi,
\qquad
\hb {\p \Psi \over \p t_{-k}} =H_{-k}\Psi,}
where we introduced the Hamiltonians
\eqn\hamil{
H_k = (L^k)_{+}+\hf (L^k)_0,
\qquad
H_{-k} =-(\bL^k)_{-}- \hf (\bL^k)_0.
}
Here the subscripts $0/-/+$ denote diagonal/lower/upper triangular 
parts of
the matrix.
From the commutativity of the second derivatives, it is easy to find
the Zakharov-Shabat zero-curvature condition
\eqn\hamil{
\hb {\p H_k\over \p t_l}-\hb {\p H_l \over \p t_k}+[H_k,H_l]=0 .
}
It means that the Hamiltonians generate commuting flows and the
perturbed system is described by the Toda Lattice hierarchy.

In addition, one can obtain the string equation for the hierarchy.
It follows from the Ward identity
\eqn\Wi{
z{\p \Psi \over \p z} ={R\over \hb}\bL^R L^R\Psi-
\left[(R+1)/2+{\alpha\over \hb}\right]\Psi
}
and can be written as
\eqn\stri{
[L^R,\bL^R] = \hb.
}
Here we should understand the operators $L^R$ as analytical 
continuation
of the operators in integer powers.

The Toda structure leads to an infinite set of PDE's for the
coefficients of the operators $L$ and $\bL$.
The first of these equations can be written for the normalization
factors and is known as Toda equation
\eqn\Todaeq{
\hb^2{\p \over \p t_1}{\p \over \p t_{-1}} \log h_n
=  {h_n\over h_{n-1}}-{h_{n+1}\over h_n} .  }
The quantity $h_n$
is known to be related to the $\tau$-function of the Toda hierarchy by
\eqn\phitau{
h_n = { \taun_{n+1} \over \taun_{n} }.
}
Taking into account \FhNP,  this gives for the partition function
\eqn\partiZ{
\ZN_{\hb}(N,t,\alpha) =\taun_{N,\hb}(t,\alpha) .
}

\appendix{B}{ Derivation of the formula for the cycle $B$}

The formula \cyclB\ can be obtained by means of the procedure of
transfer of an eigenvalue from the point $(z_1,\bz_1=\bz(z_1))$
belonging to the boundary $\gamma$ of the spot to $\infty$ (or rather
to the cut-off point $\sqrt{\L}$), worked out for the case $R=1$ in
\KM.  On the double cover described in subsection 4.3, $z_1$ and
$\bz_1$ are considered as complex coordinates of the point in
two patches. We find from \PARTFE\ that the free energy in the large
$N$ limit changes during this transfer as follows:
\eqn\FNCH{ \hb{\p\over \p N}\log\ZN=
- (z_1\bz_1)^R+V_+(z_1)+V_-(\bz_1)+\hb\sum_{m=2}^N
\log[(z_1-z_m)(\bz_1-\bz_m)], }
where all $z_m$'s are taken at their saddle point values.  Here we
took into account that the determinant in the matrix integral
\matr\ does not contribute in the quasiclassical limit.  In this limit
the integral \PARTFE\ leads to the following saddle point equations \KM
\eqn\SPEKM{ R z^R\bz^{R-1}-V_-'(\bz)=\bar G(\bz),
\qquad R z^{R-1}\bz^{R}-V_+'(z)=G(z), }
where $G(z)$ and $G(\bz)$ are the resolvents of the one dimensional
distributions of $z_k$'s and $\bz_k$'s, respectively.\foot{We view
$\bz(z)$ and $z(\bz)$ as analytical functions similarly to the
interpretation of subsection~4.3.} 
They have the following asymptotics at large $z$ and $\bz$:
\eqn\ASSZZ{ G(z)\longrightarrow {\hbar N\over z},\qquad
\bar G(\bz)\longrightarrow {\hbar N\over \bz}. }
Rewriting the sum in \FNCH\ as an integral with the measure given by
the eigenvalue density and then expressing it through the resolvents,
we find
\eqn\CONT{
 \hb {\p \over \p N}\log\ZN=-(z_1\bz_1)^R+V_+(z_1)+V_-(\bz_1) 
+ \oint {dz\over 2\pi i}\, G(z)\log(z_1-z)
+\oint {d\bz\over 2\pi i}\,\bar G(\bz)\log(\bz_1-\bz), }
where the contours of integration encircle the whole support of the
distribution of eigenvalues on the physical sheets of the functions
$\bz(z)$ and $z(\bz)$, respectively.\foot{ Note that the point of
infinite branching at the origin related to the singularities of $
z^R$ and $\bz^R$ should not appear on the physical sheets.}

Due to  \SPEKM\ we do not have
any singularities at $z\to\infty$ ($\bz\to\infty$) except
poles. Blowing up the contour we pick up the logarithmic cut and obtain
\eqn\NONL{
 \hb {\p \over \p N}\log\ZN=-(z_1\bz_1)^R-R\int_{z_1}^{\sqrt{\L}} 
{dz\over z}\, 
(z\bz(z))^R -R \int_{\bz_1}^{\sqrt{\L}} {d\bz\over \bz}\, (z(\bz)\bz)^R. }
The first term in the r.h.s. can be regrouped with the other two
terms, giving (up to a nonuniversal term $\sim\L^R$)
\eqn\ABSB{
\hb {\p \over \p N}\log\ZN=\int^{z_1}_{\sqrt{\L}} \[R(z\bz(z))^R{dz\over z} -
\hf d(z\bz)^R\] - \int_{\bz_1}^{\sqrt{\L}} \[R (z(\bz)\bz)^R{d\bz\over \bz}- 
\hf d(z\bz)^R\] }
which immediately yields \cyclBi\  if we take into account
\streq\ and the definition \chirf. Note that \SPEKM\ is essentially 
the same as \streq\ if one uses the explicit expansion \vppZ\
for the holomorphic parts of the potential. Therfore, they define the same
functions $\bz(z)$ and $z(\bz)$.   
On the complex curve described in the subsection 4.3, the formula \ABSB\
reduces to the period integral \cyclB\ of the holomorphic differential
$d\vvp$.

\listrefs 

\bye